\documentclass[reprint,aip,twocolumn,cha,showkeys,showpacs,superscriptaddress]{revtex4-1}
\usepackage[T1]{fontenc}       

\usepackage[english]{babel}

\usepackage{amsmath,amssymb,color,bm}
\usepackage{graphicx}
\usepackage{ulem}
\usepackage[caption=false]{subfig}
\usepackage{booktabs}

\renewcommand{\emph}{\textit}

\begin{document}

\title{A universal platform for magnetostriction measurements in thin films}
\date{\today}

\author{M.~Pernpeintner}
\email{matthias.pernpeintner@wmi.badw.de}
\affiliation{Walther-Mei{\ss}ner-Institut, Bayerische Akademie der Wissenschaften, D-85748 Garching, Germany}
\affiliation{Nanosystems Initiative Munich, Schellingstra{\ss}e 4, D-80799 M\"{u}nchen, Germany}
\affiliation{Physik-Department, Technische Universit\"{a}t M\"{u}nchen, D-85748 Garching, Germany}
\author{R.~B.~Holl\"ander}
\affiliation{Walther-Mei{\ss}ner-Institut, Bayerische Akademie der Wissenschaften, D-85748 Garching, Germany}
\affiliation{Physik-Department, Technische Universit\"{a}t M\"{u}nchen, D-85748 Garching, Germany}
\affiliation{Present address: Institute for Materials Science, Christian-Albrechts-Universit\"{a}t zu Kiel, D-24143 Kiel, Germany}
\author{M.\,J.~Seitner}
\affiliation{Department of Physics, University of Konstanz, D-78457 Konstanz, Germany}
\author{E.\,M.~Weig}
\affiliation{Department of Physics, University of Konstanz, D-78457 Konstanz, Germany}
\affiliation{Center for NanoScience (CeNS) and Fakult\"{a}t f\"{u}r Physik, Ludwig-Maximilians-Universit\"{a}t M\"{u}nchen, D-80799 M\"{u}nchen, Germany}
\author{R.~Gross}
\affiliation{Walther-Mei{\ss}ner-Institut, Bayerische Akademie der Wissenschaften, D-85748 Garching, Germany}
\affiliation{Nanosystems Initiative Munich, Schellingstra{\ss}e 4, D-80799 M\"{u}nchen, Germany}
\affiliation{Physik-Department, Technische Universit\"{a}t M\"{u}nchen, D-85748 Garching, Germany}
\author{S.\,T.\,B.~Goennenwein}
\affiliation{Walther-Mei{\ss}ner-Institut, Bayerische Akademie der Wissenschaften, D-85748 Garching, Germany}
\affiliation{Nanosystems Initiative Munich, Schellingstra{\ss}e 4, D-80799 M\"{u}nchen, Germany}
\affiliation{Physik-Department, Technische Universit\"{a}t M\"{u}nchen, D-85748 Garching, Germany}
\author{H.~Huebl}
\email{huebl@wmi.badw.de}
\affiliation{Walther-Mei{\ss}ner-Institut, Bayerische Akademie der Wissenschaften, D-85748 Garching, Germany}
\affiliation{Nanosystems Initiative Munich, Schellingstra{\ss}e 4, D-80799 M\"{u}nchen, Germany}
\affiliation{Physik-Department, Technische Universit\"{a}t M\"{u}nchen, D-85748 Garching, Germany}

\begin{abstract}
We present a universal nanomechanical sensing platform for the investigation of magnetostriction in thin films. It is based on a doubly-clamped silicon nitride nanobeam resonator covered with a thin magnetostrictive film. Changing the magnetization direction within the film plane by an applied magnetic field generates a magnetostrictive stress and thus changes the resonance frequency of the nanobeam. A measurement of the resulting resonance frequency shift, e.\,g.~by optical interferometry, allows to quantitatively determine the magnetostriction constants of the thin film. We use this method to determine the magnetostriction constants of a 10\,nm thick polycrystalline cobalt film, showing very good agreement with literature values. The presented technique can be useful in particular for the precise measurement of magnetostriction in a variety of (conducting and insulating) thin films, which can be deposited by e.\,g.~electron beam deposition, thermal evaporation or sputtering.
\end{abstract}

\maketitle
Nanomechanical systems are an established platform for mass and force detection. In particular, the high quality factors of their vibrational modes\cite{unterreithmeier_damping_2010} make them ideally suited for high-precision sensing applications in (nano)biology, medicine, chemistry and physics\cite{degen_nanoscale_2008,eom_nanomechanical_2011,boisen_cantilever-like_2011,arlett_comparative_2012,moser_ultrasensitive_2013}. For example, nanomechanical resonators allow for the detection of DNA molecules\cite{mukhopadhyay_nanomechanical_2005} and atoms\cite{chaste_nanomechanical_2012}, and nanomechanical resonance spectroscopy has been proposed as a versatile tool and extension of conventional spectroscopy techniques in biology and chemistry\cite{greaney_nanomechanical_2008}. In solid state physics, nanomechanical sensors are utilized for  the investigation of material properties of thin films\cite{schmid_damping_2011,karabalin_stress-induced_2012,faust_signatures_2014}, which often significantly differ from those of bulk materials\cite{sorensen_magnetic_1924,richter_mechanical_2010}. One particular aspect is the investigation of externally tunable material properties as discussed in the field of multiferroics\cite{spaldin_renaissance_2005}. For example, it has been demonstrated that magnetostriction and magnetic anisotropy in a Ga$_{0.948}$Mn$_{0.052}$As thin film can precisely be investigated using a nanomechanical beam setup\cite{masmanidis_nanomechanical_2005}.

Here, we extend this concept and present a universal platform for the experimental investigation of magnetostrictive thin films which uses a doubly-clamped silicon nitride (Si$_3$N$_4$) nanobeam covered with a thin layer of the material of interest. This approach allows for the investigation of any magnetostrictive thin film material---conducting as well as insulating---which can be deposited on a Si$_3$N$_4${} nanobeam, using e.\,g.~electron beam evaporation, thermal evaporation or sputtering.
As the thin film deposition is the last step in the sample fabrication process, the ferromagnetic film is not exposed to etching solution or dry etch reactants, which allows to apply this technique to a broad range of materials. Moreover, the measurement sensitivity is expected to be independent of the film thickness which could be useful for the investigation of very thin magnetostrictive films.

\begin{figure}[tb]
\centering
\includegraphics[scale=1]{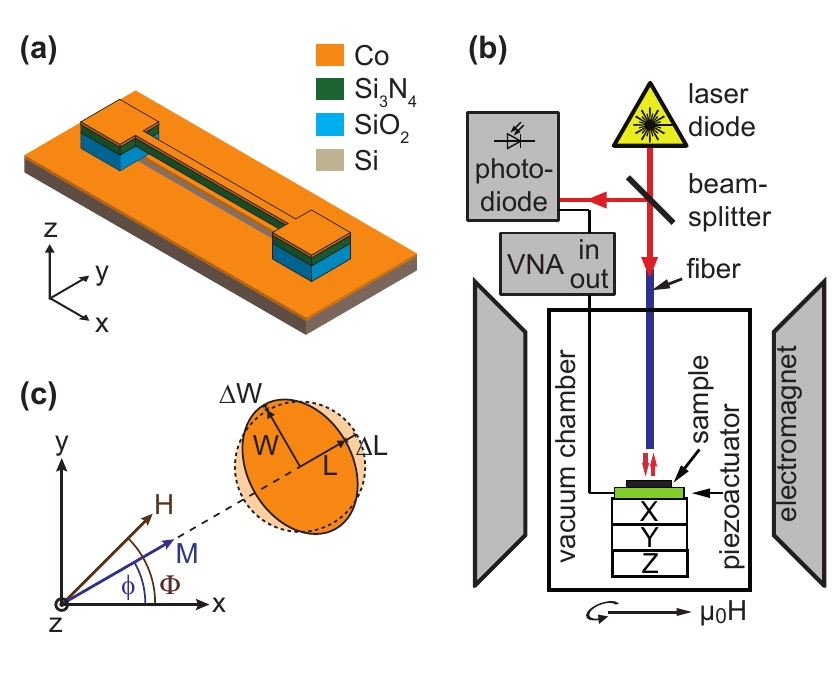}
\caption{\textbf{(a)} Schematic of the doubly-clamped Si$_3$N$_4${} nanobeam (green) covered with a thin cobalt film (orange). \textbf{(b)} Fiber interferometer used to characterize the sample in vacuum. \textbf{(c)} Coordinate system with the nanobeam along the $x$-axis. The external magnetic field and the magnetization in the cobalt film are oriented in the $x$-$y$-plane with an angle $\Phi$ resp.~$\phi$ to the $x$-axis. A free-standing cobalt film (circularly shaped for illustration purposes) is deformed by magnetostriction along and perpendicular to the magnetization axis.}
\label{fig:ExperimentalSetupAndCoordinateSystem}
\end{figure}

We start the sample fabrication with a single-crystalline silicon wafer commercially coated with a $200\,\mathrm{nm}$ thick thermal oxide and a $t_{\mathrm{SiN}}=90\,\mathrm{nm}$ thick LPCVD (low pressure chemical vapor deposition) high-stress Si$_3$N$_4${} film. We use electron beam lithography, aluminium evaporation and lift-off to create an etch mask for the $l=25\,\mathrm{\mu m}$ long and $350\,\mathrm{nm}$ wide, doubly-clamped nanomechanical beam. With a reactive ion etching step, we pattern transfer the structure to the silicon nitride layer. We subsequently remove the aluminium mask and release the nanobeam with buffered hydrofluoric acid. Finally, we deposit $t_{\mathrm{film}}=10\,\mathrm{nm}$ of cobalt on the chip using electron beam evaporation. Figure~\ref{fig:ExperimentalSetupAndCoordinateSystem}a shows a schematic of the Si$_3$N$_4$/Co bilayer nanobeam.

To spectroscopically investigate the sample, we use the fiber interferometer (laser wavelength: $673\,\mathrm{nm}$) sketched in Fig.~\ref{fig:ExperimentalSetupAndCoordinateSystem}b (see also Ref.~\citenum{vogel_optically_2003}) operating at room temperature and in vacuum ($p<10^{-4}\,\text{mbar}$) to avoid air damping. An electromagnet provides a homogeneous magnetic field at the sample position. In this setup, any beam displacement is translated linearly into a change in the detected photovoltage, which allows to investigate the mechanical motion of the nanobeam\cite{azak_nanomechanical_2007}. Using a piezoelectric actuator we can resonantly drive the flexural out-of-plane motion of the beam. Employing signal vector analysis we study the mechanical response of the beam as function of the applied actuator drive frequency. Additionally, to control the magnetization direction of the nanobeam, we use the static magnetic field provided by the electromagnet, which we are able to rotate in the sample plane (i.\,e.~the $x$-$y$-plane). The azimuthal angle $\Phi$ is defined between the external field direction and the main beam axis (see Fig.~\ref{fig:ExperimentalSetupAndCoordinateSystem}a,\,c).

We next present a model predicting the add-on stress on the magnetic layer caused by the magnetostrictive effect. For this, we start with a free-standing polycrystalline magnetic film in an external magnetic field. We further assume the external field to be sufficiently large to fully magnetize the ferromagnet (i.\,e.~align all magnetic moments). Due to magnetostriction, the magnetic film is mechanically deformed along (perpendicular to) the magnetization direction $\mathbf{M}/M_{\mathrm{s}}$ as sketched in Fig.~\ref{fig:ExperimentalSetupAndCoordinateSystem}c. The relative contraction/elongation for the directions along and perpendicular to the magnetization orientation is given by the magnetostrictive constants $\lambda_{\parallel}=\Delta L/L$ and $\lambda_{\perp}=\Delta W/W$, respectively. For cobalt (as well as iron or nickel), $\lambda_{\parallel}<0$ and $\lambda_{\perp}>0$. In the coordinate system introduced in Fig.~\ref{fig:ExperimentalSetupAndCoordinateSystem}, the resulting strain vector reads\cite{chikazumi_physics_1997} (see Supporting Information I)
\begin{equation*}
\boldsymbol\epsilon_{\mathrm{mag}}=\begin{pmatrix}\epsilon_{xx}\\ \epsilon_{yy}\\ \epsilon_{zz}\end{pmatrix}=\begin{pmatrix}\lambda_{\parallel}\cos^2(\phi)+\lambda_{\perp}\sin^2(\phi)\\ \lambda_{\parallel}\sin^2(\phi)+\lambda_{\perp}\cos^2(\phi)\\ \lambda_{\perp} \end{pmatrix}.
\end{equation*}
Here, $\phi$ denotes the angle between the nanobeam axis and the magnetization orientation in the ferromagnetic film, as defined in Fig.~\ref{fig:ExperimentalSetupAndCoordinateSystem}c.

In case of a thin magnetic film deposited on a substrate, the shared interface imposes a boundary condition on the magnetic film. Thus, instead of strain, a magnetostrictive stress $\boldsymbol\sigma_{\mathrm{mag}}=-\mathbf{C}\cdot \boldsymbol\epsilon_{\mathrm{mag}}$ arises in the film (see Supporting Information I), where $\mathbf{C}$ is the stiffness tensor of the magnetic material. For a polycrystalline material, as studied here, $\mathbf{C}$ can be expressed in terms of Young's modulus $E$ and shear modulus $\mu$ (see, e.\,g., Refs.~\citenum{gross_festkoerperphysik_buch,weiler_voltage_2009}). The stress $\boldsymbol\sigma_{\mathrm{mag}}$ in the magnetic film is then given by (see Supporting Information I)
\begin{equation*}
\boldsymbol\sigma_{\mathrm{mag}}=-\begin{pmatrix}E\lambda_{\parallel}\cos^2(\phi)\\ E\lambda_{\parallel}\sin^2(\phi)\\ 0 \end{pmatrix},
\end{equation*}
where the magnetostriction constants are related by $\lambda_{\perp}=-\nu\lambda_{\parallel}$ as volume magnetostriction can be neglected in first order\cite{chikazumi_physics_1997}. Here, $\nu$ denotes the Poisson ratio of the magnetostrictive material, $\nu=E/(2\mu)-1$ (see Ref.~\citenum{hunklinger_festkoerperphysik_2009}).

To obtain the effective stress present in the double-layer beam, we take into account the stress present in the silicon nitride, $\sigma_{\mathrm{SiN}}$, as well as in the magnetic thin film, $\boldsymbol\sigma_{\mathrm{film}}^\mathrm{tot}$. For a double-layer system, the effective stress along the beam direction is given by \cite{hocke_determination_2014,seitner_damping_2014}
\begin{equation*}
\sigma_{\mathrm{eff}}=\frac{\sigma_{\mathrm{SiN}} t_{\mathrm{SiN}} + \sigma_{\mathrm{film},x}^\mathrm{tot}t_{\mathrm{film}}}{t_{\mathrm{SiN}}+t_{\mathrm{film}}}.
\end{equation*}
Note that the total stress in the magnetic layer contains stresses from the fabrication process as well as the magnetostrictive stress, $\boldsymbol\sigma_{\mathrm{film}}^\mathrm{tot}=\boldsymbol\sigma_{\mathrm{film}}^0+\boldsymbol\sigma_{\mathrm{mag}}$.

For a highly tensile-stressed nanobeam, the resonance frequency of the fundamental flexural mode is well approximated by $\omega_{\mathrm{res}}/2\pi=1/(2l)\sqrt{\sigma/\rho}$, where $\sigma$ and $\rho$ denote pre-stress and density of the nanobeam (see e.\,g.~Ref.~\citenum{verbridge_high_2006}). Using effective values for stress and density of the double-layer beam\cite{hocke_determination_2014,seitner_damping_2014}, we obtain
\begin{equation}\label{eq:resfreqFull}
\frac{\omega_{\mathrm{res}}(\phi)}{2\pi}=\frac{1}{2l}\sqrt{\frac{\sigma_{\mathrm{eff}}}{\rho_{\mathrm{eff}}}}=\frac{1}{2l}\sqrt{\frac{\sigma_0-\sigma_1\cos^2(\phi)}{\rho_{\mathrm{eff}}}},
\end{equation}
where we have defined $\sigma_0=(\sigma_{\mathrm{SiN}} t_{\mathrm{SiN}}+\sigma_{\mathrm{film},x}^0 t_{\mathrm{film}})/(t_{\mathrm{SiN}}+t_{\mathrm{film}})$ and $\sigma_1=E t_{\mathrm{film}}\lambda_{\parallel}/(t_{\mathrm{SiN}}+t_{\mathrm{film}})$.

For $\sigma_1/\sigma_0\ll 1$, we can Taylor-expand Eq.~(\ref{eq:resfreqFull}) and obtain for the relative resonance frequency shift
\begin{equation}\label{eq:resfreqShift}
\frac{\Delta\omega_{\mathrm{res}}(\phi)}{\omega_{\mathrm{res,}0}} = \frac{\omega_{\mathrm{res}}(\phi)-\omega_{\mathrm{res,}0}}{\omega_{\mathrm{res,}0}} = -\frac{\sigma_1}{2\sigma_0}\cos^2(\phi)
\end{equation}
with $\omega_{\mathrm{res},0}/2\pi=1/(2l)\sqrt{\sigma_0/\rho_{\mathrm{eff}}}$ the resonance frequency at $\phi=90^{\circ}$.

We thus expect a $\cos^2(\phi)$-dependence of the resonance frequency as a function of the magnetization direction. By measuring the tuning amplitude $\sigma_1/2\sigma_0$ of the resonance frequency, we can deduce the magnetostriction constants $\lambda_{\parallel}$ and $\lambda_{\perp}$.

To experimentally investigate such a magnetostrictive tuning of a nanomechanical beam, we measure the resonance frequency of the beam in an external magnetic field of constant magnitude $\mu_0H=200\,\mathrm{mT}$, which is above the coercive field and the in-plane anisotropy fields of the Co film. In this regime, the magnetization is, in a good approximation, aligned along the external field direction ($\phi=\Phi$). We rotate the external field between $\Phi=-35^{\circ}$ and $200^{\circ}$, where $\Phi=0^{\circ}$ corresponds to $\mu_0\mathbf{H} \parallel \mathbf{\hat{x}}$ (see Fig.~\ref{fig:ExperimentalSetupAndCoordinateSystem}c).

\begin{figure}[tb]
\centering
\includegraphics[scale=1]{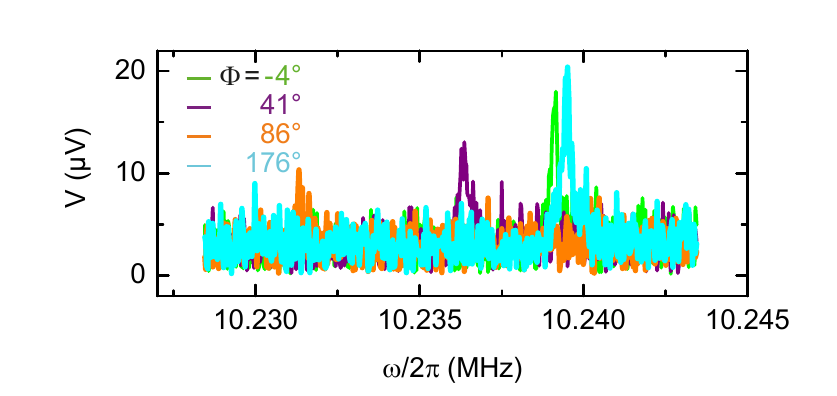}
\caption{Measured photovoltage spectra for an external magnetic field of $\mu_0H=200\,\mathrm{mT}$ applied in different directions $\Phi=-4^{\circ},41^{\circ},86^{\circ},176^{\circ}$.}
\label{fig:AmplitudeSpectra}
\end{figure}

Figure \ref{fig:AmplitudeSpectra} shows the measured homodyne photovoltage as a function of the drive frequency for four different external field directions $\Phi=-4^{\circ}, 41^{\circ}, 86^{\circ}, 176^{\circ}$. We observe a clear resonance peak corresponding to the fundamental vibrational out-of-plane mode of the nanobeam, whose frequency shifts as a function of the magnetic field orientation. The magnetostrictive frequency shift significantly exceeds the linewidth of the resonance peaks, which is $\Gamma_{\mathrm{m}}/2\pi\approx 300\,\mathrm{Hz}$. In addition to the resonance frequency shift, we observe a variation of the resonance peak amplitude when rotating the external magnetic field vector, as one can see in Fig.~\ref{fig:AmplitudeSpectra}.
We attribute this to a slight translational shift of the sample position in high magnetic fields. In this case the laser spot is not centered on the nanobeam any more, which decreases the detected photovoltage amplitude.

\begin{figure}[tb]
\centering
\includegraphics[scale=1]{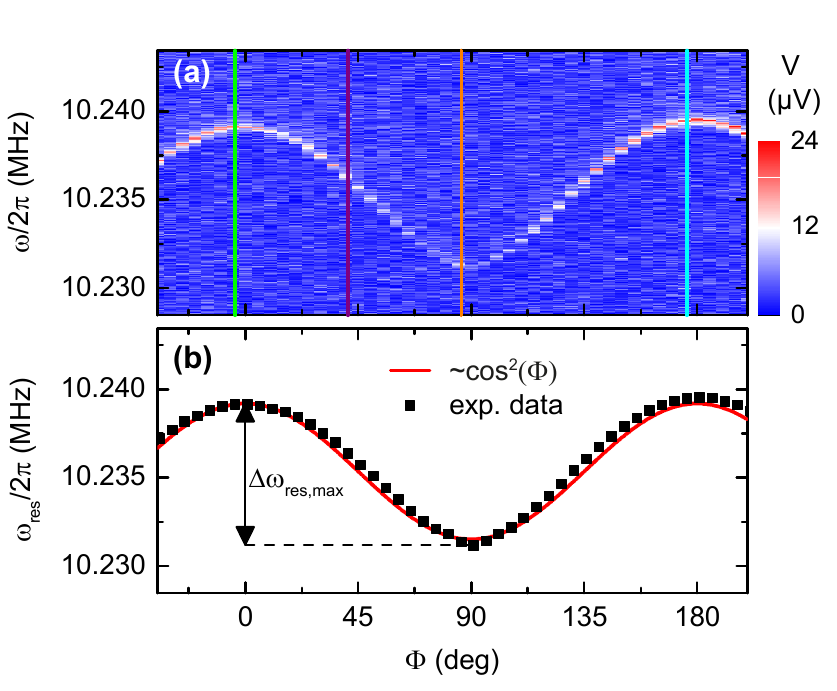}
\caption{Resonance frequency of the fundamental flexural mode as a function of the external magnetic field orientation $\Phi$ for $\mu_0H=200\,\mathrm{mT}$. \textbf{(a)} shows the measured photovoltage $V$ as a function of drive frequency and external field direction. In \textbf{(b)}, the fitted resonance frequency $\omega_{\mathrm{res}}(\Phi)$ is compared to the expected $\cos^2(\Phi)$ behaviour.}
\label{fig:ResonanceFrequencyAngleScan}
\end{figure}

To analyze the observed field orientation dependence in more detail, we measure the amplitude spectrum of the nanobeam as a function of $\Phi$, as shown in Fig.~\ref{fig:ResonanceFrequencyAngleScan}a. The experimental data confirm the expected $180^{\circ}$-periodicity of the resonance frequency $\omega_{\mathrm{res}}(\Phi)$. From these data we extract the maximum resonance frequency shift $\Delta\omega_{\mathrm{res,max}}:=\omega_{\mathrm{res}}(0^{\circ})-\omega_{\mathrm{res}}(90^{\circ})=2\pi\times 8.00\,\mathrm{kHz}$ (see Fig.~\ref{fig:ResonanceFrequencyAngleScan}b). We observe slight deviations between the measured $\omega_{\mathrm{res}}(\Phi)$ and the expected $\cos^2(\Phi)$ behaviour, especially around $\Phi=45^{\circ}$ and $135^{\circ}$. This is quantitatively understood as the magnetization for these angles $\Phi$ is not perfectly aligned in parallel with the applied magnetic field due to the shape anisotropy of the nanobeam (for details see Supporting Information II).

To determine the magnetostriction constants $\lambda_{\parallel}$ and $\lambda_{\perp}$ from the experimental data, we first calculate the static prestress $\sigma_0$ in the nanobeam. With $\omega_{\mathrm{res}}(\phi=90^{\circ})/2\pi=1/(2l)\sqrt{\sigma_0/\rho_{\mathrm{eff}}}$ we get $\sigma_0=892\,\mathrm{MPa}$. Here we have used the effective density $\rho_{\mathrm{eff}}=3410\,\mathrm{kgm^{-3}}$, which we determine from $\rho_{\mathrm{SiN}}=2800\,\mathrm{kgm^{-3}}$ (see Ref.~\citenum{unterreithmeier_damping_2010}) and $\rho_{\mathrm{Co}}=8900\,\mathrm{kgm^{-3}}$ (see Ref.~\citenum{greenwood_chemie_1988}) using $\rho_{\mathrm{eff}}=(\rho_{\mathrm{SiN}}t_{\mathrm{SiN}}+\rho_{\mathrm{Co}}t_{\mathrm{film}})/(t_{\mathrm{SiN}}+t_{\mathrm{film}})$ (see Ref.~\citenum{hocke_determination_2014}).
With Eq.~(\ref{eq:resfreqShift}), the Young's modulus of cobalt $E=175\,\mathrm{GPa}$ (see Ref.~\citenum{masumoto_thermal_1967}) and the relation $\lambda_{\perp}=-\nu\lambda_{\parallel}$, the observed maximum resonance frequency shift of $\Delta\omega_{\mathrm{res,max}}=2\pi\times 8.00\,\mathrm{kHz}$ corresponds to a magnetostriction constant of $\lambda_{\parallel}=-79.7\times 10^{-6}$ ($\lambda_{\perp}=27.9\times 10^{-6}$).

For comparison with literature, we calculate the magnetostriction constants $\lambda_{\parallel}$ and $\lambda_{\perp}$ in polycrystalline magnetically saturated cobalt from the crystal magnetostriction constants $\lambda_{\text{A,B,C,D}}$ as described in Ref.~\citenum{mason_derivation_1954}, for which we obtain $\lambda_{\parallel}=-78.4\times 10^{-6}$ and $\lambda_{\perp}=27.5\times 10^{-6}$. This is in very good agreement with the experimentally determined values. Thus we conclude that the proposed method conforms with standard magnetostriction measurement techniques, which e.\,g.~use the bending of a cantilever covered with a thin magnetostrictive film\cite{klokholm_measurement_1976,tam_precise_1988,du_tremolet_de_lacheisserie_magnetostriction_1994,weber_uhv_1994}, and is therefore suitable to quantitatively determine the magnetostriction constants of thin films.
In contrast to cantilever-based experiments, where magnetostriction causes a bending of the mechanical element, the present approach uses a pre-stressed, doubly-clamped nanobeam where the magnetostrictive stress modifies the total stress along the beam axis and therefore changes the resonance frequency of the beam.
This stress-to-frequency conversion allows for an effective determination of the magnetostriction constants via a frequency measurement, which does not rely on a quantitative measurement of the beam displacement (as it is the case for cantilever-based techniques).
The high quality factor of pre-stressed Si$_3$N$_4$ nanobeam resonators\cite{unterreithmeier_damping_2010} therefore allows to precisely investigate magnetostriction in thin films.

Although the presented method only allows to experimentally access one stress direction (i.\,e.~the stress component along the beam direction), it can be particularly useful for the investigation of very thin (or nanopatterned) magnetostrictive films with a high precision. The reason is that, as we will show in the following, the experimental uncertainty of the calculated magnetostriction constants does not necessarily increase for decreasing film thickness as it is the case for cantilever-based measurement techniques.

To illustrate this, we first calculate the stress-frequency gauge factor, i.\,e.~the change of the resonance frequency of the beam as a function of the stress variation. For the presented sample this is
\begin{equation*}
\frac{\Delta\omega_{\mathrm{res}}}{\sigma_{\mathrm{mag},x}}=-\frac{\omega_{\mathrm{res},0}t_{\mathrm{film}}}{2\sigma_0 (t_{\mathrm{SiN}}+t_{\mathrm{film}})}=2\pi\times 0.57\,\mathrm{Hz/kPa}.
\end{equation*}

Assuming a frequency measurement precision of $\delta\omega_{\mathrm{res}}\approx\Gamma_{\mathrm{m}}/2$ (with the linewidth of the resonance $\Gamma_{\mathrm{m}}$), this allows to resolve a stress variation of $\delta\sigma_{\mathrm{mag},x}=0.52\,\mathrm{MPa}$. This corresponds to an experimental uncertainty in the parallel magnetostriction constant of $\delta\lambda_{\parallel}=\delta\sigma_{\mathrm{mag},x}/E=3.0\times 10^{-6}$. Using a phase-locked loop (PPL) to track the resonance frequency of the beam, however, would increase the frequency resolution significantly, allowing an uncertainty $\delta\lambda_{\parallel}$ well below $10^{-6}$. This is comparable to other methods\cite{klokholm_measurement_1976,tam_precise_1988,weber_uhv_1994,masmanidis_nanomechanical_2005} even though the thickness of our magnetostrictive film is only 10\,nm. In particular, reducing the film thickness further does not necessarily reduce the measurement precision. This is due to the fact that the quality factor of a doubly-clamped Si$_3$N$_4${} beam covered with a thin film typically strongly depends on the film thickness. For Si$_3$N$_4$/Au nanobeams, it has been shown recently that the inverse quality factor is proportional to the film thickness for Au layers between 10\,nm and 100\,nm as the damping in a highly prestressed silicon nitride film is much lower than in the Au film\cite{seitner_damping_2014}. Therefore, for very thin magnetostrictive films on a highly prestressed Si$_3$N$_4${} beam, the resonance frequency measurement uncertainty is proportional to the film thickness, $\delta\omega_{\mathrm{res}}\propto t_{\mathrm{film}}$. The uncertainty $\delta\lambda_{\parallel}\propto \delta\sigma_{\mathrm{mag},x}=2\delta\omega_{\mathrm{res}} \sigma_0(t_{\mathrm{film}}+t_{\mathrm{SiN}})/(\omega_{\mathrm{res},0}t_{\mathrm{film}})$ is therefore in first approximation independent of the film thickness (assuming $t_{\mathrm{film}}\ll t_{\mathrm{SiN}}$). This characteristic makes the proposed technique an ideal platform for the investigation of magnetostriction in thin and ultrathin films.

In this letter, we have proposed a new method to quantitatively investigate magnetostriction in thin films. To this end we use a Si$_3$N$_4${} nanomechanical resonator covered with a thin magnetostrictive film. By measuring the resonance frequency of the fundamental vibrational mode of the beam as a function of an external magnetic field we can deduce the magnetostrictive stress along the beam direction and hence the magnetostriction constants $\lambda_{\parallel}$ and $\lambda_{\perp}$. Compared to previously reported methods, the proposed technique does not rely on a quantitative measurement of the mechanical displacement, but utilizes a resonance frequency shift caused by magnetostriction. Besides, it offers a measurement precision which is independent of the film thickness. This enables the investigation of ultrathin magnetostrictive films and paves the way to study magnetostriction as a function of the film thickness. The proposed technique can be applied to any conducting or insulating material which can be deposited on a Si$_3$N$_4${} nanobeam via electron beam evaporation, thermal evaporation, sputtering etc. The material under investigation does not have to be etch-resistent as it is deposited on the nano-resonator as last step of the sample fabrication process.

The authors thank Dr.~S.~Gepraegs for fruitful discussions and gratefully acknowledge financial support by the Deutsche Forschungsgemeinschaft via Project No.~Ko 416/18 and SpinCAT GO944/4.

%

\pagebreak
\widetext
\vspace{1cm}
\begin{center}
\textbf{\large Supporting Information: A universal platform for magnetostriction measurements in thin films}
\end{center}

\renewcommand{\thetable}{\Roman{table}}
\renewcommand{\thesection}{\Roman{section}}
\renewcommand*{\citenumfont}[1]{S#1}
\renewcommand*{\bibnumfmt}[1]{[S#1]}

\makeatletter
\makeatother

\renewcommand{\thesection}{\Alph{section}}
\renewcommand{\thesubsection}{\alph{subsection}}
\renewcommand{\thefigure}{S\arabic{figure}}
\renewcommand{\thetable}{S\arabic{table}}
\renewcommand{\theequation}{S\arabic{equation}}
\renewcommand{\refname}{Additional References}

\setcounter{equation}{0}


\section{Derivation of the magnetostrictive stress $\boldsymbol\sigma_{\mathrm{mag}}$}
In this section, we derive the magnetostrictive stress vector $\boldsymbol\sigma_{\mathrm{mag}}$. We start with the magnetostrictive deformation of a polycristalline free-standing ferromagnetic (FM) film with cylindrical shape and transform the corresponding strain tensor to the $xyz$-coordinate system (where the $x$-axis is parallel to the nanobeam). Then we use this result to calculate the magnetostrictive stress in a FM thin film on a substrate.

\begin{figure}[htb]
\centering
\includegraphics[scale=1]{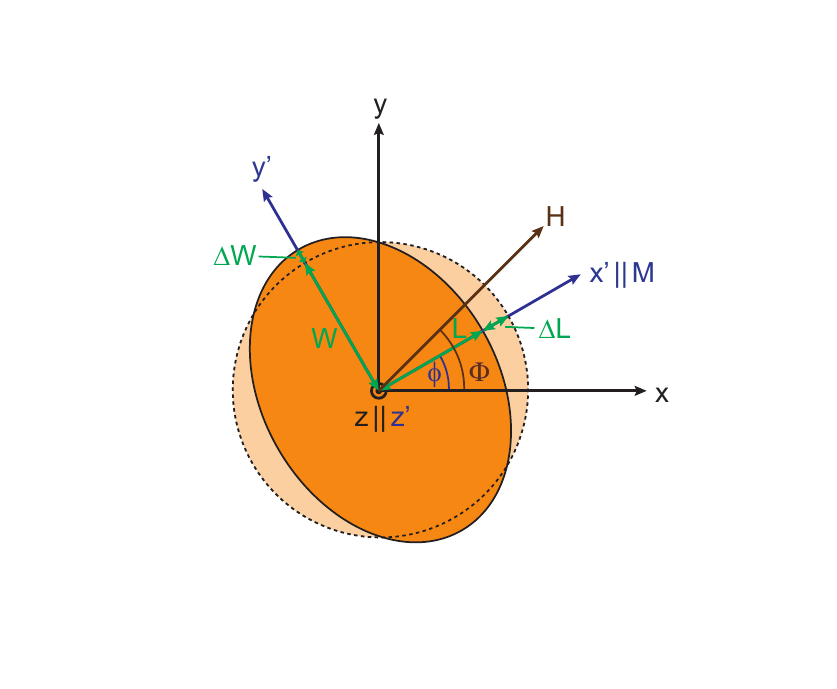}
\caption{Definition of the $x'y'z'$ coordinate system which is rotated by $\phi$ with respect to the $xyz$ coordinate system, with $z'\parallel z$. The magnetization is along the $x'$-axis, while the nanobeam is parallel to the $x$-axis. Aligning the magnetization in a free-standing magnetostrictive thin film (for illustration purposes we have chosen a cylindrically shaped film here) causes a strain $\lambda_{\parallel}$ ($\lambda_{\perp}$) along the $x'$ ($y'$ and $z'$) axis.}
\label{fig:CoordinateSystemAndMagnetostrictionSM}
\end{figure}

First, we consider a free-standing FM thin film, which is centered around the origin of the $xyz$-coordinate system (the lab system), as illustrated in Fig.~\ref{fig:CoordinateSystemAndMagnetostrictionSM} (dashed circle). When magnetizing the FM, the film is contracted (elongated) along (perpendicular to) the magnetization direction. The relative contraction/elongation $\epsilon_{\parallel,\perp}$ is given by\cite{si_chikazumi_physics_1997}
\begin{equation*}
\epsilon_{\parallel}=\frac{\Delta L}{L}=\lambda_{\parallel}\quad\text{and}\quad
\epsilon_{\perp}=\frac{\Delta W}{W}=\lambda_{\perp}.
\end{equation*}

We define a second coordinate system $x'y'z'$, which is rotated relative to the $xyz$-system by an angle of $\phi$ around the $z$-axis. Here, $\phi$ is the angle between the nanobeam axis and the magnetization direction as shown in Fig.~\ref{fig:CoordinateSystemAndMagnetostrictionSM}. The $x'y'z'$-coordinate system represents the natural system for the magnetostriction. In this frame of reference, the magnetization $\mathbf{M}$ is along the $x'$-axis and the magnetostrictive strain tensor is given by
\begin{equation*}
\boldsymbol\epsilon'_{\text{mag}}=\begin{pmatrix}\lambda_{\parallel} & 0 & 0\\ 0 & \lambda_{\perp} & 0\\ 0 & 0 & \lambda_{\perp}\end{pmatrix}.
\end{equation*}

Using the rotation tensor $\mathbf{R}$, which maps the $xyz$-coordinate system to the $x'y'z'$-coordinate system,
\begin{equation*}
\mathbf{R} = \begin{pmatrix}\cos\phi & -\sin\phi & 0\\ \sin\phi & \cos\phi & 0\\ 0 & 0 & 1\end{pmatrix},
\end{equation*}
we get the components of the strain tensor in the $xyz$-coordinate system of the nanobeam:
\begin{equation*}
\boldsymbol\epsilon_{\text{mag}} = \mathbf{R}^T\boldsymbol\epsilon'_{\text{mag}}\mathbf{R} =
    \begin{pmatrix}\lambda_{\parallel}\cos^2\phi + \lambda_{\perp}\sin^2\phi & (\lambda_{\perp}-\lambda_{\parallel})\cos\phi\sin\phi & 0\\ (\lambda_{\perp}-\lambda_{\parallel})\cos\phi\sin\phi & \lambda_{\perp}\cos^2\phi + \lambda_{\parallel}\sin^2\phi & 0 \\ 0 & 0 & \lambda_{\perp}\end{pmatrix}
\end{equation*}

In case of a thin FM film on a substrate (or on top of a Si$_3$N$_4${} nanobeam, as in our case), the geometric dimensions of the FM are fixed in $x$- and $y$-direction, which means the effective strain along these axes vanishes. We describe this with an additional strain $\epsilon_{\text{boundary}}$ so that the net strain
\begin{equation*}
\boldsymbol\epsilon_{\text{net}} = \boldsymbol\epsilon_{\text{mag}}+\boldsymbol\epsilon_{\text{boundary}}
\end{equation*}
vanishes along the $x$- and $y$-direction:
\begin{equation*}
\epsilon_{\text{net},x} = \epsilon_{\text{net},y} = 0.
\end{equation*}
In a more intuitive picture, magnetostriction changes the equilibrium dimensions of the film. Along the magnetization direction, e.g., the equilibrium length of the film is reduced ($\lambda_{\parallel} < 0$). However, the boundary conditions require an unchanged length, which results in a tensile stress in the FM film along the $x$-axis. Analogously, magnetostriction creates a compressive stress in the FM along the $y$-direction. Perpendicular to the film plane, the FM is free to expand.

Thus, the strain applied by the boundary conditions creates the stress
\begin{equation}\label{sigmamag}
\boldsymbol\sigma_{\text{mag}} = \mathbf{C} \boldsymbol\epsilon_{\text{boundary}}
\end{equation}
with the elasticity tensor $\mathbf{C}$ (see Ref.~\citenum{si_gross_festkoerperphysik_buch}).

In Voigt notation (see e.\,g.~Ref.~\citenum{si_gross_festkoerperphysik_buch}), $\epsilon_{\text{boundary}}$ is written as
\begin{equation}\label{epsboundary}
\mathbf{\epsilon_{\text{boundary}}} = \begin{pmatrix}\epsilon_{\text{boundary},xx}\\ \epsilon_{\text{boundary},yy}\\ \epsilon_{\text{boundary},zz}\\ 2\epsilon_{\text{boundary},yz}\\ 2\epsilon_{\text{boundary},xz}\\ 2\epsilon_{\text{boundary},xy}\end{pmatrix}
    = \begin{pmatrix}\lambda_{\parallel}\cos^2\phi + \lambda_{\perp}\sin^2\phi \\ \lambda_{\perp}\cos^2\phi + \lambda_{\parallel}\sin^2\phi\\ \epsilon_{\text{boundary},zz}\\ 0\\ 0\\ 2(\lambda_{\perp}-\lambda_{\parallel})\cos\phi\sin\phi\end{pmatrix}
\end{equation}
and the elasticity tensor $\mathbf{C}$ is given by
\begin{equation}\label{Ctensor}
\mathbf{C} = \begin{pmatrix}\lambda+2\mu & \lambda & \lambda & 0 & 0 & 0\\
    \lambda & \lambda+2\mu & \lambda & 0 & 0 & 0\\
    \lambda & \lambda & \lambda+2\mu & 0 & 0 & 0\\
    0 & 0 & 0 & \mu & 0 & 0\\
    0 & 0 & 0 & 0 & \mu & 0\\
    0 & 0 & 0 & 0 & 0 & \mu \end{pmatrix}
\end{equation}
with the shear modulus $\mu$, the Lam\'{e} constant $\lambda=(2\mu^2-E\mu)/(E-3\mu)$ and Young's modulus $E$ (see Refs.~\citenum{si_hunklinger_festkoerperphysik_2009,si_gross_festkoerperphysik_buch}).

As the FM film can expand freely in $z$-direction, the stress component $\sigma_{\text{mag},zz}$ has to vanish. This results in
\begin{equation*}
\epsilon_{\text{boundary},zz}=\lambda_{\parallel}\left(1-\frac{E}{2\mu}\right),
\end{equation*}
where we have used $\lambda_{\parallel}=-\nu\lambda_{\perp}$ (see main text) and $\mu=E/(2(1+\nu))$ (see Ref.~\citenum{si_gross_festkoerperphysik_buch}).

With this, Eqs.~(\ref{sigmamag}--\ref{Ctensor}) lead to
\begin{equation*}
\sigma_{\text{mag},xx}=-E\lambda_{\parallel}\cos^2(\phi)
\end{equation*}
and
\begin{equation*}
\sigma_{\text{mag},yy}=-E\lambda_{\parallel}\sin^2(\phi).
\end{equation*}

\section{Impact of imperfect magnetization alignment on the resonance frequency shift}
As mentioned in the main text, the measured $\Phi$-dependence of the nanobeam resonance frequency shows slight deviations from the expected $\cos^2(\Phi)$ behaviour. To study this in more detail, we repeat the measurements discussed in the main article for lower magnetic fields. Whereas for $\mu_0H=200\,\mathrm{mT}$ the magnetization $\mathbf{M}$ is roughly aligned along $\mu_0\mathbf{H}$, this is not the case for smaller external fields where the magnetic anisotropy becomes an issue.

\begin{figure}[tbh]
\centering
\includegraphics[scale=1]{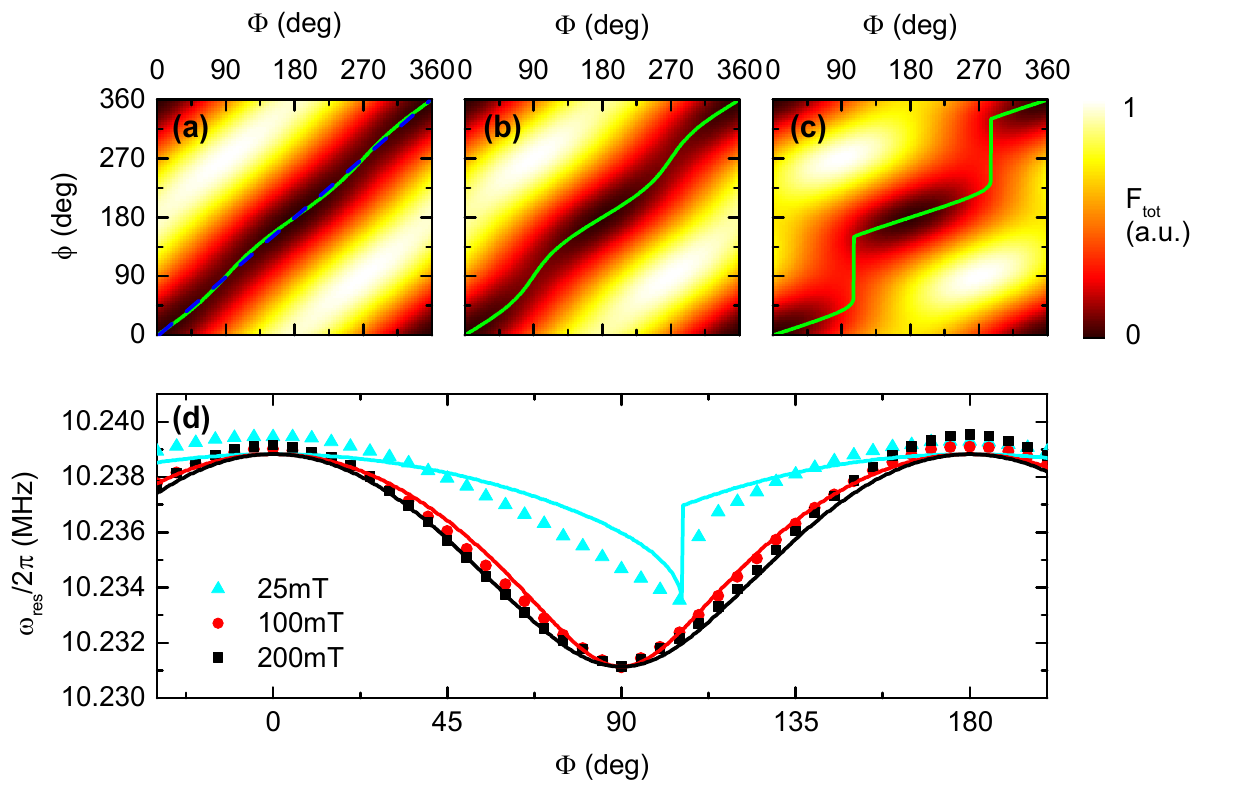}
\caption{\textbf{(a-c)} Total free energy density as a function of external magnetic field orientation $\Phi$ and magnetization direction $\phi$ for $\mu_0H=200\,\mathrm{mT}$ \textbf{(a)}, $100\,\mathrm{mT}$ \textbf{(b)} and $25\,\mathrm{mT}$ \textbf{(c)}. The green lines indicate the energy density minima. The colorbar has been rescaled individually for each graph. \textbf{(d)} Resonance frequency versus external field orientation $\Phi$ for various field strengths $\mu_0H=200,100,25\,\mathrm{mT}$. The symbols represent the experimental data, the modelled resonance frequency is plotted as solid lines.}
\label{fig:AngleScanVariousFields}
\end{figure}

To model this, we start with a Stoner-Wolfarth approach and assume a saturated magnetization state $|\mathbf{M}|=M_{\mathrm{s}}$ (see e.\,g.~Ref.~\citenum{si_chikazumi_physics_1997}). To determine the equilibrium direction of the magnetization as a function of the external field, we write down the free energy density $F_{\mathrm{tot}}$, which is the sum of Zeeman energy density and shape anisotropy. The first is given by $F_{\mathrm{Zeeman}}=-\mu_0\mathbf{M}\cdot\mathbf{H}$ (see Ref.~\citenum{si_chikazumi_physics_1997}), the latter is $F_{\mathrm{aniso}}=(\mu_0/2)\mathbf{M}\cdot\mathbf{N}\cdot\mathbf{M}$, where $\mathbf{N}$ denotes the demagnetization tensor\cite{si_morrish_physical_2001}. We approximate the cobalt thin film on the Si$_3$N$_4${} nanobeam as an ellipsoid with axis lengths $l$, $w$ (width of the nanobeam) and $t_{\text{film}}$. The demagnetization tensor is then diagonal with the components $N_{xx}\simeq 0$, $N_{yy}=0.03$ and $N_{zz}=0.97$ (see Ref.~\citenum{si_brandlmaier_ma_2006}). Note that we here neglect crystalline anisotropy contributions to $F_{\text{tot}}$, assuming that they average out in a polycrystalline film\cite{si_weiler_voltage_2009}. The contribution of magnetoelastic energy to $F_{\text{tot}}$ is about two orders of magnitude smaller than the contribution of shape anisotropy and can therefore be neglected.

For the present experimental geometry, external magnetic field and magnetization are in the $x$-$y$-plane (see Fig.~\ref{fig:CoordinateSystemAndMagnetostrictionSM}). Thus, the total free energy density is given by
\begin{equation}
F_{\mathrm{tot}}=-\mu_0M_{\mathrm{s}} H\cos(\Phi-\phi) + \frac{\mu_0M_{\mathrm{s}}^2}{2}N_{yy}\sin^2(\phi).
\end{equation}

In Fig.~\ref{fig:AngleScanVariousFields}a-c, the total free energy density is plotted as a function of the external magnetic field orientation and the magnetization direction for the parameter values $M_{\mathrm{s}}=1167\,\mathrm{kA/m}$ (measured by SQUID magnetometry for a similar Co thin film), $N_{yy}=0.03$ and $\mu_0H=200\,\mathrm{mT}$, $100\,\mathrm{mT}$ and $25\,\mathrm{mT}$. The minima, which determine the equilibrium magnetization direction, are highlighted by green lines. If the external field significantly exceeds the demagnetization field $\mu_0H_{\mathrm{demag}}=\mu_0M_{\mathrm{s}} N_{yy}\approx22\,\mathrm{mT}$, the magnetization is approximately parallel to the external field, i.\,e.~$\phi\approx\Phi$. 
Even for $\mu_0H=200\,\mathrm{mT}$, however, there are slight deviations between the calculated energy density minima and the ideal case $\Phi=\phi$ (indicated with the dashed blue line in Fig.~\ref{fig:AngleScanVariousFields}a). 
For small external fields $\mu_0H\lesssim\mu_0H_{\mathrm{demag}}$, the shape anisotropy of the thin cobalt stripe forces the magnetization direction towards the $x$-axis, as Fig.~\ref{fig:AngleScanVariousFields}c points out.


In Fig.~\ref{fig:AngleScanVariousFields}d, the measured resonance frequency is plotted as a function of the external field orientation $\Phi$ for the field strengths $\mu_0H=200\,\mathrm{mT}$,$100\,\mathrm{mT}$ and $25\,\mathrm{mT}$. The solid lines show the modelled resonance frequency behaviour, based on Eq.~(2) (main text) and the calculated equilibrium magnetization orientation $\phi(\Phi)$. In particular for $\mu_0H=200\,\mathrm{mT}$, we observe excellent agreement between the experimental data and the model. For small magnetic fields, e.\,g.~$\mu_0H=25\,\mathrm{mT}$, our simple model explains the resonance frequency shift at least qualitatively, reproducing the switching of the magnetization at $\Phi=110^{\circ}$ and the measured maximum frequency shift.

\end{document}